\documentclass[11pt]{JHEP3}

\usepackage{amssymb,amsmath}
\usepackage[dvips]{graphicx}
\usepackage{epsfig}
\usepackage{amsfonts}

\usepackage{amscd}

\def\be{\begin{eqnarray}}
\def\ee{\end{eqnarray}}

\def\e{\epsilon}

\def\d{\partial}

\newcommand{\idn}{{1\relax{\kern-.35em}1}}

\newcommand{\Rf}{\mathbb{R}}

\newcommand{\hO}{{\hat O}}

\def\e{\epsilon}

\preprint{ WIS/18/07-OCT-DPP }

\title{The space-time operator product expansion in string theory duals of field theories}
\author{Ofer Aharony and Zohar Komargodski\\
Department of Particle Physics, The Weizmann Institute of Science\\
Rehovot 76100, Israel\\ \email{Ofer.Aharony@weizmann.ac.il},
\email{zkomargo@wisemail.weizmann.ac.il}}

\abstract{We study the operator product expansion (OPE) limit of correlation functions in field theories which
possess string theory duals, from the point of view of the string worldsheet. We show how the interesting
(``single-trace'') terms in the OPE of the field theory arise in this limit from the OPE of the worldsheet
theory of the string dual, using a dominant saddle point which appears in computations of worldsheet correlation
functions in the space-time OPE limit. The worldsheet OPE generically contains only non-physical operators, but
all the non-physical contributions are resummed by the saddle point to a contribution similar to that of a
physical operator, which exactly matches the field theory expectations. We verify that the OPE limit of the
worldsheet theory does not have any other contributions to the OPE limit of space-time correlation functions.
Our discussion is completely general and applies to any local field theory (conformal at high energies)
that has a
weakly coupled string theory dual (with arbitrary curvature). As a first application, we compare our results to
a proposal of R.~Gopakumar for the string theory dual of free gauge theories.}

\begin{document}

\section{Introduction and summary}

When some field theory (such as a large $N$ gauge theory) has a string theory dual \cite{'t
Hooft:1973jz,adscft}, this dual should exhibit all aspects of the field theory. In particular, a basic feature
of any local field theory is the existence of an operator product expansion (OPE) when two operators approach
each other (in Euclidean space), expressing the product of two nearby operators in terms of a sum of local
operators, and it is interesting to ask how this is realized in the string theory dual. We will focus here on
specific terms in this OPE, which are ``single-trace'' terms in the language of large $N$ gauge theories. We
will see that
these terms arise from the OPE limit of the
worldsheet correlation functions of the dual string theory. In fact, each term corresponding to some operator in
the space-time OPE expansion is reproduced by a saddle point in the worldsheet computation, which occurs when
the distance between the two operators on the worldsheet is proportional to the space-time distance between the
two operators, raised to some power (depending on the operator). Our discussion will be limited to tree-level
closed string theory, but it applies to any field theory which is dual to a closed string theory in some
background (with arbitrary curvature). It would be interesting to generalize our discussion to higher genus
(namely, higher orders in $1/N$ when the field theory is a large $N$ gauge theory). It would also be interesting
(and presumably straightforward) to generalize our results to theories with open strings.

Of course, in most of the interesting cases of string theory duals of four dimensional large $N$ gauge theories,
such as the large $N$ limit of QCD, we do not yet know the worldsheet action well enough to analyze it in
detail. However, it turns out that general properties, following just from the existence of such a worldsheet,
are sufficient to give a precise worldsheet description of the space-time OPE. Cases where the worldsheet theory
is under control, for instance string theory on $AdS_3\times \mathcal W$ with NS-NS 3-form flux, and string
theory on weakly curved $AdS_{d+1}\times W$ spaces with Ramond-Ramond flux, arise as special cases of our
general description. In the $AdS_3$ case we can make contact with previously known results
\cite{Maldacena:2001km}, and obtain a new perspective on these results.

For simplicity, we will analyze only the case of conformal field
theories, which are dual to string theories on anti-de Sitter (AdS)
spaces; since the OPE is a short-distance property, this analysis
should be relevant to any local field theory which is conformal at high energies.
It would be interesting
to generalize our results to asymptotically free
gauge theories, where there are logarithmic corrections to the OPE. Our
analysis seems to depend strongly on the fact that the string theory
duals of conformal field theories have a continuum of off-shell modes,
and it would be interesting to understand if this is true also for the string
duals of asymptotically free theories or not \cite{Brower:2006ea}.

We should clarify the relation of this work to previous works on similar issues. There are many computations of
correlation functions in field theories dual to string theory on $AdS$ space (in particular, $AdS_5\times S^5$)
in the supergravity approximation. These computations allow the study of the OPE limit of these supergravity
correlators, and its matching with the field theory expectations; for example, see \cite{SUGRA-GAUGE OPE} and
references therein. There is no direct relation between our results and these studies; we focus on the
worldsheet description rather than the bulk space-time description. Of course, in the special cases where the
string theory has a supergravity approximation, our results are consistent with the bulk space-time
computations, but our point of view is completely different.

The saddle point that we find in the integral over the moduli space
in the computation of the worldsheet correlation functions is somewhat
similar to the saddle point found in \cite{Gross:1987ar} for high-energy
scattering. However, in that case the saddle point is in the
bulk of the worldsheet moduli space (even in the extreme high-energy
limit), while in our case the position of the saddle
point scales with the small separation of the operators in the
field theory. Since our saddle point is associated with an OPE, it
depends only on the properties of the two operators involved and
not on any other operators, while the saddle point of \cite{Gross:1987ar}
depends on all the vertex operators.

Finally, in the series of papers \cite{Polchinski-Strassler,Polchinski:2002jw,Brower:2006ea}, a systematic study
was performed of various kinematical limits in gauge theories, relating them with the behavior in the dual
string theory. Our study is very much in the same spirit and has, in particular, a close relation to the work
\cite{Polchinski:2002jw} on deep inelastic scattering.\footnote{In the physical region of deep inelastic
scattering (DIS) the momentum fraction $x$ satisfies $0\leq x\leq 1$, which is not the OPE region. However, OPE
techniques are very useful in analyzing DIS, since they provide moments of the structure functions. This follows
by a standard contour argument connecting the strict OPE limit ($x\gg 1$) with the physical region.} However,
the methods we use are quite different.

This paper is organized as follows. In \S 2 we show how a worldsheet computation (using the worldsheet OPE)
reproduces the expected terms in the space-time OPE (the ones involving ``single-trace'' operators). We show
that these terms arise from a saddle point, and that there are no perturbative corrections to the saddle point
approximation. In \S 3 we analyze various discrete contributions which appear in the computation, and we show
that in general they are not associated with the space-time OPE. As an application of our results, in \S 4 we
compare our general results to the OPE limit in a proposal of \cite{Gopakumar:2005fx} for the string theory dual
of free gauge theories.

\section{The string theory dual of the space-time OPE} \label{main}

\subsection{The general problem}

Consider the Euclidean background of string theory on
$AdS_{d+1}\times W$, which is related by the AdS/CFT
correspondence \cite{adscft} to a conformal field theory (CFT) in
$d$ dimensions.\footnote{Here and elsewhere, we use the notation
of a direct product although our analysis includes also
cases where the $AdS_{d+1}$ is non-trivially fibered over the base
$W$. } The conformal theory lives on $\Rf^d$ or on $S^d$ (the two
cases are related by a conformal transformation), and we will
denote its coordinates by $\vec{x}$. The worldsheet theory
describing strings on $AdS_{d+1}\times W$ is not tractable in
the general case when Ramond-Ramond (RR) backgrounds are turned on,
but it certainly exists, for example, in the
Green-Schwarz or pure spinor formalisms.

The basic observables in the space-time CFT are correlation functions of local operators $\hO_i(\vec{x})$. Each
such operator sits in some representation of the space-time conformal algebra $SO(d+1,1)$. For simplicity we
will discuss here only scalar operators which are primaries of the conformal algebra; the generalization to
other operators is straightforward. Such operators are characterized by a space-time scaling dimension, which we
will denote by $2j$ (We reserve the notation $\Delta$ for worldsheet scaling dimensions; this convention agrees
with the standard convention for strings in $AdS_3$.), and in general also by additional quantum numbers, which
we will denote collectively by $h$. In CFTs which are dual to weakly coupled string theories (with arbitrary
curvature), there is a special class of operators which create states which are dual to single-string states;
for large $N$ gauge theories these are the ``single-trace'' operators (and we will use this name for them also
more generally). For simplicity, we will limit our discussion to such operators. The duality maps these
operators to integrated vertex operators on the worldsheet (We will discuss
 here only tree-level diagrams for which the worldsheet is a sphere.).
 This is written schematically as
\begin{equation}\label{mapping}
\hO_{j,h}(\vec{x})=\int d^2z\, O_{j,h}(\vec{x};z,\bar z).
\end{equation}
The OPE of two ``single-trace'' operators contains both other ``single-trace'' operators and ``multi-trace''
operators. On general grounds involving the form of the 't Hooft expansion \cite{'t Hooft:1973jz} we would
expect the ``single-trace'' terms in the space-time OPE to arise locally on the worldsheet, from the limit where
the two operators approach each other also in their worldsheet ($z$) coordinates, while other terms in the
space-time OPE should not arise locally.

The space-time OPE of conformal primary operators is highly
constrained by the conformal group $SO(d+1,1)$ : it takes the
form, as $\vec{x}' \to \vec{x}$,
\begin{equation}\label{stOPE}
    \hO_{j_1,h_1}(\vec{x}') \hO_{j_2,h_2}(\vec{x})\sim \sum_{j,h} \frac{C(j_1,j_2,j;h_1,h_2,h)}{(|\vec{x}'-\vec{x}|^2)^{(j_1+j_2-j)}} \hO_{j,h}(\vec{x})
    + \cdots,
\end{equation}
where the dots denote descendants of the space-time conformal algebra. The coefficients $C$ are determined by
the 3-point functions and the 2-point functions. Using the OPE, one can relate a limit of $n$-point functions in
the space-time CFT (with $n > 3$) to $(n-1)$-point functions. To make the notations less cluttered, we will
replace the quantum numbers $\{j_i,h_i\}$ simply by $i$.

\begin{figure}[tbp]
\begin{center}
\includegraphics{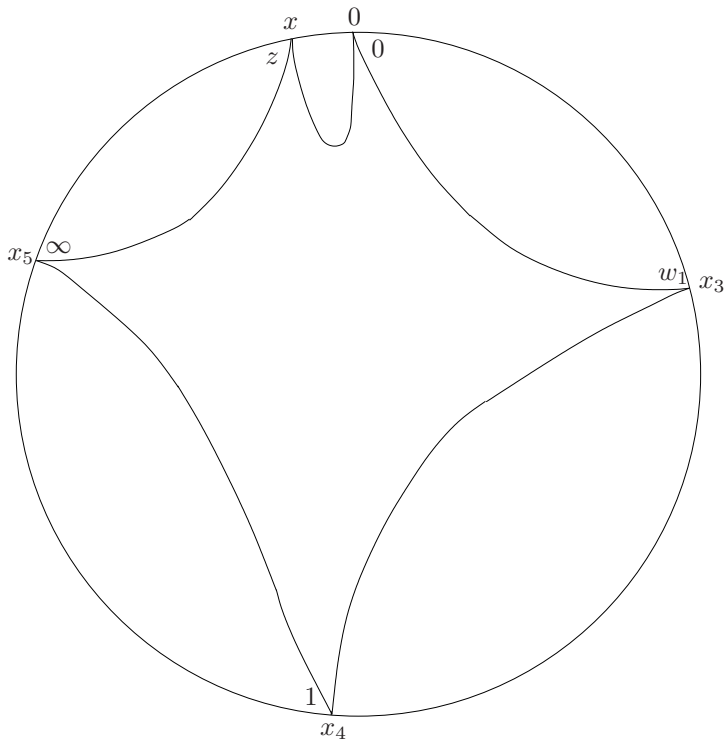} \caption{Two space-time operators are separated by $x$, which is taken to be very small
compared to the other space-time distances. The corresponding worldsheet insertions are separated by the modulus
$z$, which is integrated over. The figure is a schematic drawing of the string worldsheets contributing to the
Euclidean 5-point function \protect\eqref{thegeneralproblem}. The result we will find is that for each term in
the OPE the moduli space integral is dominated by a saddle point with $z$ proportional to $x$ raised to some
positive power. Hence, $z$ is small, in accordance with our geometrical intuition.} \label{worldsheetspacetime}
\end{center}
\end{figure}

The field theory/string theory duality allows us to compute $n$-point functions in the space-time CFT via
$n$-point functions of integrated vertex operators on the corresponding worldsheet. For convenience, we can fix
one of the operators in space-time to be at $\vec{x}=0$.\footnote{We can do more using the conformal symmetry,
but we will not need this.} We will take one of the other operators to be at $\vec{x}$, and study the limit
$\vec{x}\to 0$. The $n$-point function in the space-time CFT (with $n > 3$) is given by
\begin{gather}\label{thegeneralproblem}
    <\hO_1(\vec{x})\hO_2(0)\prod_{i=1}^{n-4}\hO_{i+2}(\vec{x}_{i+2})\hO_{n-1}(\vec{x}_{n-1})\hO_{n}(\vec{x}_{n})>=
    \cr=
    \int d^2z\left<O_{1}(\vec{x};z,\bar{z})\bar ccO_{2}(0;0)\left(\prod_{i=1}^{n-4} \int d^2w_iO_{i+2}(\vec{x}_{i+2};w_i)\right)
    \bar ccO_{n-1}(\vec{x}_{n-1};1)\bar ccO_{n}(\vec{x}_n;\infty)\right>\equiv\cr\equiv
    \int d^2z\left<O_{1}(\vec{x};z,\bar{z})\bar ccO_{2}(0;0)X\right>,
\end{gather}
where $X$ is an abbreviation for all the other operators; we integrate over all possible worldsheets with the
correct insertions. The worldsheet distance between the two insertions which correspond to the space-time
operators which are separated by $\vec{x}$ is denoted by $z$, see figure \ref{worldsheetspacetime}. We have used
the worldsheet conformal symmetry to fix three of the vertex operators at $z=0,1,\infty$, and the usual ghost
insertions take care of the ghost number anomaly. The prescription (\ref{thegeneralproblem}) computes only the
connected planar diagrams in the space-time CFT.

The worldsheet vertex operators appearing in \eqref{mapping} are
always physical vertex operators; they are level-matched Virasoro
primaries with total worldsheet scaling dimension $\Delta=2$.
However, the worldsheet OPE of such operators generally contains
unphysical operators on the worldsheet, corresponding to
``off-shell modes'' in AdS space (just like in the worldsheet OPE
of string theory in flat space). In AdS space, fields may be taken
off-shell by changing their momentum in the radial direction, or,
equivalently, by changing the representation of the space-time
conformal algebra which they sit in. These representations are
labeled by the space-time scaling dimension $j$. The space-time
operators discussed above sit in representations of the conformal
algebra with $j \geq d/4$, \footnote{The case of $(d-2)/4 \leq j <
d/4$ can also be dealt with but it is more complicated
\cite{Breitenlohner:1982jf,Klebanov:1999tb}.} which map to
non-normalizable modes in AdS space. The other unitary
representations of the space-time conformal algebra have $j = d/4
+ i s$ with real $s$, and correspond to (delta-function)
normalizable modes in AdS, with some radial momentum. In general
we expect to find in the worldsheet OPE ``off-shell modes'' which
have a form similar to the physical vertex operators, but with
other values of $j$ which correspond to unitary representations.
We will denote these off-shell modes by the same notation
$O_{j,h}$, noting that for given quantum numbers $h$ the operator
is only physical for a specific value of $j$, and we will denote
their worldsheet scaling dimension by
\begin{equation}
    \Delta(O_{j,h})\equiv 2\Delta(j,h).
\end{equation}
The existence
of such off-shell modes (with any value of $j$) is obvious when the
worldsheet theory factorizes
as a direct product of conformal field theories $AdS_{d+1}\times W$, but
we expect them to exist also in other cases (such as $AdS_5\times S^5$).

Note that the Hilbert space of the worldsheet theory includes only the (delta-function) normalizable vertex
operators,\footnote{This is a general property of holographic backgrounds; the local operators in space-time do
not correspond to normalizable states of the worldsheet theory. This phenomenon was noted already in Liouville
theory \cite{Seiberg:1990eb}, and another example is $AdS_3$
\cite{Teschner:1997ft,Teschner:1997fv,Giveon:1998ns,deBoer:1998pp,Kutasov:1999xu,Teschner:1999ug}.} with $j =
d/4 + is$, so naively one would think that only these operators appear in the worldsheet OPE. In particular, the
vertex operators on the worldsheet which correspond to local operators in space-time through the mapping
(\ref{mapping}) are not part of the Hilbert space of the worldsheet theory. On the other hand, these
non-normalizable operators should appear somehow in the worldsheet OPE in order to reproduce the space-time OPE
\eqref{stOPE}. This is the main puzzle we would like to solve, by analyzing the space-time OPE limit of
\eqref{thegeneralproblem} (More precisely, we will analyze the contribution to this limit from the worldsheet
OPE region.).

Before we begin, let us give a couple of examples of our
general statements. String theory on $AdS_3$ with NS-NS 3-form
flux is described for any value of the curvature by a WZW coset
model $SL(2,\mathbb C)/SU(2)$. In this case, the scaling
dimensions and the spectrum of primary operators are known exactly
\cite{Teschner:1997ft,Teschner:1997fv,Giveon:1998ns,deBoer:1998pp,Kutasov:1999xu,
Teschner:1999ug}, and are given by
\begin{equation}\label{ads3nsns}\Delta(j,h)=\frac{j(1-j)}{k-2}+h_0,\end{equation}
where $k$ is the level of $SL(2)$ (For large $k$ this is the same as the radius squared of $AdS_3$ in string
units.) and $h_0$ is the scaling dimension of the operator in the other components of the worldsheet CFT.

Another tractable case is weakly curved $AdS_{d+1}\times W$ backgrounds.
As long as these backgrounds are weakly
curved, the worldsheet dimensions of scalar primary operators are given by the
space-time Laplacian (in units of the string scale). Hence, for this case,
\begin{equation} \label{largevol}
\Delta(j,h)=\frac{j(d/2-j)}{R_{AdS}^2}+h_0,
\end{equation}
where $h_0$ is now the contribution from the Laplacian on $W$.

Note that the worldsheet scaling dimensions are real for all values of $j$ corresponding to unitary
representations of the conformal group. In both examples described above the worldsheet scaling dimensions
become increasingly negative for large $j$ (going to $-\infty$ as $j\to +\infty$), and increasingly positive for
large $|s|$ in $j=d/4+is$ (going to $+\infty$ as $|s|\to \infty$), and their first and second derivatives with
respect to $j$ are negative for any $j
> d/4$. We will assume that these properties are true also more generally, for any string theory on AdS space,
since (as we will see below) they are necessary in order to reproduce correctly the space-time OPE.

\subsection{The worldsheet OPE and its relation to the space-time OPE}
\label{ope_comp}

In general we do not know how to compute
\eqref{thegeneralproblem}, but we can compute it in the limit of
small $|z|$ using the worldsheet OPE; as discussed above, this
limit is expected to give the ``single-trace'' terms in the
space-time OPE. In a general string theory on AdS space it is
not known how to directly compute the worldsheet OPE of the
non-normalizable vertex operators that we are interested in.
However, the worldsheet OPE of normalizable vertex operators (with
$j=d/4+is$) is expected to include only other normalizable vertex
operators (which span the Hilbert space of the worldsheet CFT),
with coefficients determined by the 3-point functions. The
worldsheet OPE thus includes an integral over the contour
$j=d/4+is$, which we will denote by $C$. We can then determine the
OPE for the non-normalizable vertex operators by (careful)
analytic continuation in $j$; this is the standard procedure for
computing correlation functions in string theory on $AdS_3$, and
we expect it to be valid more generally. Note that the worldsheet
does not in general factorize into decoupled left and right moving
sectors due to the RR fields. In addition, it does not necessarily
split into a sum of two decoupled conformal field theories
associated to $AdS$ and to $W$ (again due to the RR fields and
possible fibration).

For normalizable vertex operators with $j_1,j_2\in C$, the most general form of the worldsheet OPE is,
as $z\to 0$,
\begin{equation}\label{general OPE}
    O_{1}(\vec{x};z)O_{2}(0;0)=\sum_h\int_C dj \int d^d x' F(z;\vec{x},\vec{x}';j_i,j;h_i,h)
    O_{j,h}(\vec{x}';0)+descendants,
\end{equation}
where $F$ is related to the 2-point and 3-point functions on the
worldsheet. For other values of $j_1$ and $j_2$, which are relevant for the
correlators of space-time CFT operators that we are interested in,
the formula (\ref{general OPE}) needs to be analytically
continued in $j_1$ and $j_2$; this continuation yields additional discrete
contributions from poles of $F$ which we will discuss later, but otherwise
it preserves the form of \eqref{general OPE}. We assume that the
function $F$ is analytic in the $j$'s except at these poles; this is true in all known examples,
and we expect it to always be true.

Conformal invariance fixes completely the $z$ and $x$ dependence of $F$. This results in (ignoring the
worldsheet descendants and denoting $|y|=\sqrt{\vec{y}\cdot \vec{y}}$)
\begin{equation}\label{OPE1}
    O_{1}(\vec{x};z)O_{2}(0;0)=\sum_h\int_C dj \int d^d x' \frac{|z|^{2(-\Delta(1)-\Delta(2)+\Delta(j,h))}}{|x|^\alpha|x'|^\beta|x'-x|^\gamma}
    F(j_i,j;h_i,h) O_{j,h}(\vec{x}';0),\end{equation}
where the powers $\alpha$, $\beta$ and $\gamma$ (depending on $j_1$, $j_2$ and $j$) may be fixed using the
space-time conformal invariance. Denoting $\vec{x}=|x|\hat x$ we can rewrite the integral as
\begin{equation}\label{manipulatedintegral2}
    O_{1}(\vec{x};z)O_{2}(0;0)=\sum_h\int_C dj\frac1{|x|^{\alpha+\beta+\gamma}} \int d^d x' \frac{|z|^{2(-\Delta(1)-\Delta(2)+\Delta(j,h))}}
    {(|x'|/|x|)^\beta(|x'/|x|-\hat x|)^\gamma}
    F O_{j,h}(\vec{x}';0),\end{equation}
    where the arguments of $F$ are suppressed.
Changing variables to the dimensionless ratio $\vec{y}=\vec{x}'/|x|$ we obtain
\begin{equation}\label{manipulatedintegral3}
    O_{1}(\vec{x};z)O_{2}(0;0)=|z|^{-2\Delta(1)-2\Delta(2)}\sum_h\int_C dj\frac{|z|^{2\Delta(j,h)} F}{|x|^{\alpha+\beta+\gamma-d}}
    \int d^d y \frac{1}
    {|y|^\beta|\vec{y}-\hat x|^\gamma}
    O_{j,h}(\vec{y}|x|;0).\end{equation}
Since we are interested in the small $|x|$ limit and we are ignoring the subleading contributions from
space-time descendants, we can expand the last operator for small $x$ and keep only the first term in its Taylor
expansion,
\begin{equation}\label{manipulatedintegral3.5}
    O_{1}(\vec{x};z)O_{2}(0;0)=\sum_h \int_C dj\frac{|z|^{-2\Delta(1)-2\Delta(2)}}{|x|^{\alpha+\beta+\gamma-d}} |z|^{2\Delta(j,h)}FO_{j,h}(0;0)\int d^d y \frac{1}
    {|y|^\beta|\vec{y}-\hat x|^\gamma}.
     \end{equation}
     The final integral is now simply a number (depending on the $j$'s but independent of $\hat x$); the same number appears
     in the relation of $F$ to the 3-point function which appears in the space-time OPE.
Next, acting with a space-time dilation on \eqref{OPE1} gives
\begin{equation}\alpha+\beta+\gamma-d=2j_1+2j_2-2j,\end{equation}
while the fact that the two operators whose OPE we are interested in
are physical operators leads to
\begin{equation}-\Delta(1)-\Delta(2)=-2.\end{equation}
Using these results we obtain the
final form of the OPE in the limit we are interested in,
\begin{equation}\label{manipulatedintegral4}
    O_{1}(\vec{x};z)O_{2}(0;0)=\sum_h \int_C dj\frac{|z|^{-2}}{|x|^{2j_1+2j_2-2j}} |z|^{2\Delta(j,h)-2}FO_{j,h}(0;0)\int d^d y \frac{1}
    {|y|^\beta|\vec{y}-\hat x|^\gamma}.
     \end{equation}
This can be plugged back into our general $n$-point function
(\ref{thegeneralproblem}), giving
\begin{gather}\label{final-result}
    I \equiv \int d^2z\left<O_{1}(\vec{x};z,\bar{z})\bar ccO_{2}(0;0)X\right>=\cr=\sum_h \int d^2z\int_C dj\frac{|z|^{-2}}{|x|^{2j_1+2j_2-2j}} |z|^{2\Delta(j,h)-2}F\left<\bar cc O_{j,h}(0;0) X
    \right>\int d^d y \frac{1}
    {|y|^\beta|\vec{y}-\hat x|^\gamma}=\cr=\frac{2\pi}{|x|^{2j_1+2j_2}}\sum_h \int d\ln |z|\int_C dj|z|^{2\Delta(j,h)-2}|x|^{2j} F\left<\bar cc O_{j,h}(0;0) X
    \right>\int d^d y \frac{1}
    {|y|^\beta|\vec{y}-\hat x|^\gamma}.
\end{gather}
The expression above is not exact, but it accounts for the leading
contribution to the $z$ integral from the OPE region (in space-time
and on the worldsheet). Since all
the contributions from Virasoro descendants have been dropped, this is a
good approximation of the $z$ integral in some small region
around $z=0$, say $|z|<\epsilon$.

We will now argue that in the space-time OPE limit of $|x|\to 0$, the double integral in \eqref{final-result} is
dominated (for each value of $h$) by a saddle point which reproduces the ``single-trace'' terms in the
space-time OPE. To see this more easily, we rewrite (\ref{final-result}) as
\begin{gather}\label{theintegralthree}
   I = |x|^{-2j_1-2j_2} \sum_h\int d\ln{|z|}\int_{C} dj|z|^{2(\Delta(j,h)-1)}|x|^{2j} B(j_i,h_i;j,h),
\end{gather}
where $B$ stands for the product of all the factors which are independent of $x$ and $z$ : $2\pi F$, the
integral over $y$, and the $(n-1)$-point correlator of the worldsheet theory. In the limit of small $|x|$, the
integral over $j$ (for each $h$) has a saddle point at $j=j_0$ such that
\begin{equation}\label{j saddle}
    \d_j\Delta(j,h)|_{j=j_0} =-\frac{\ln |x|}{\ln |z|}.
\end{equation}

\begin{figure}[tbp]
\begin{center}
\includegraphics{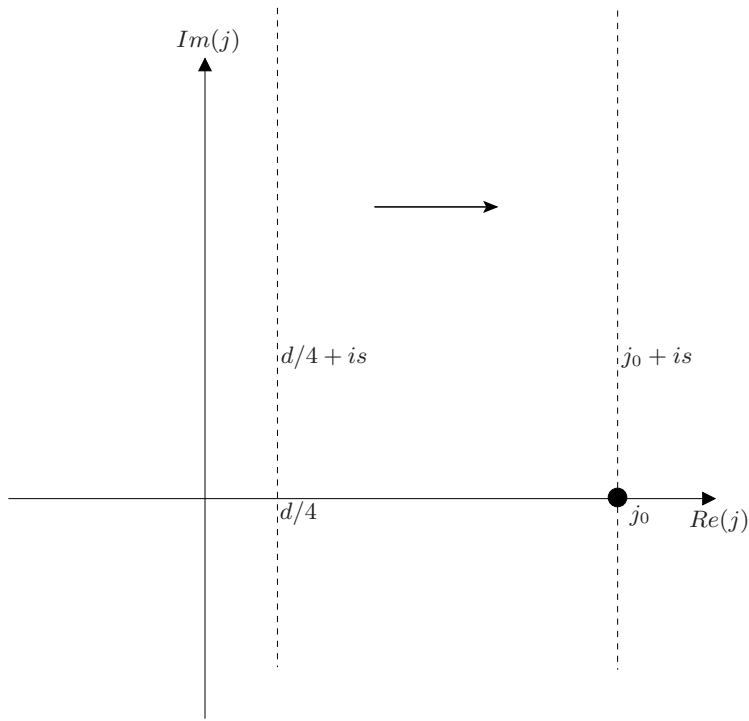} \caption{We can shift the contour from $d/4+is$ to $j_0+is$ since the integrand is
meromorphic (Poles crossed as a result of this shift are analyzed in \S 3.). The saddle point is on the real
axis and is denoted by the bold circle at $j=j_0$.} \label{contourshift}
\end{center}
\end{figure}

We will see below that for the dominant values of $z$ the saddle point \eqref{j saddle} has $j_0\geq d/4$. This
is generally off the integration contour, but we expect the integrand to be an analytic function of $j$ (with
specific poles), so we can shift the integration contour $C$ in the direction of the positive real axis so that
it intersects the real axis at $j_0$ (see figure \ref{contourshift}). \footnote{The function $B$ has various
poles, and we may get some discrete contributions when we cross them as we move the contour. We will analyze
these in the next section. Our assumption that $\Delta\to +\infty$ as $|Im(j)|\to +\infty$ for fixed $Re(j)$
implies that the integral over finite horizontal lines at infinity vanishes (This is needed to apply the Cauchy
theorem.).} The saddle point is then dominant if
\begin{equation}\label{saddle point existence condition}
\ln|z| \cdot \d_j^2\Delta(j,h)|_{j=j_0} \gg 0.\end{equation}

We will see that the dominant contributions come from the region of small $z$ where this is arbitrarily large
(in the $x\to 0$ limit), so that the saddle point approximation is justified in this limit (using our assumption
that $\d_j^2 \Delta < 0$). Performing the $j$ integral along this new contour at leading order in the saddle
point approximation we obtain
\begin{gather}\label{z integral}
    I = i\sqrt{\pi}|x|^{-2j_1-2j_2} \sum_h\int \frac{d\ln|z|}{\sqrt{\ln|z|\d_j^2\Delta|_{j_0}}}|x|^{2j_0}e^{2\ln|z|\left(\Delta(j_0,h)-1\right)}B(j_i,h_i;j_0,h).
\end{gather}
This is still a complicated integral since $j_0$ generally depends in an intricate way on $\ln|z|$ (through
\eqref{j saddle}). We can make progress by using the saddle point method again for the integral in \eqref{z
integral}. But first, it is convenient to change variables from $\ln|z|$ to the ratio $w \equiv \ln |z| / \ln
|x|$. The integral becomes
\begin{gather}\label{y integral}
    I = -i\sqrt{\pi}|x|^{-2j_1-2j_2} \sum_h\int \frac{\sqrt{-\ln |x|} dw}{\sqrt{-w\d_j^2\Delta|_{j_0}}}|x|^{2j_0}e^{2\ln|x|w\left(\Delta(j_0,h)-1\right)}B(j_i,h_i;j_0,h),
\end{gather}
where the overall minus sign is because of the change in the lower
and upper limits of the integral. Up to an overall factor of $2$,
the exponent in the integral is
\begin{equation}\label{exponent}
    \ln|x| \left[ w(\Delta(j_0,h)-1)+j_0\right].
\end{equation}
The first derivative with respect to $w$ gives, upon using (\ref{j saddle}),
\begin{equation}
\ln|x| \frac{\d}{\d w}\left(w\Delta(j_0,h)-w+j_0\right)=\ln |x|(\Delta(j_0,h)-1).
\end{equation}
Hence, the location of the saddle point in the integral over $w$ (or $z$) is determined by the solution of the
equation
\begin{equation}\label{z saddle}
    \Delta(j_0,h)=1.
\end{equation}
This is precisely the condition for a non-normalizable physical vertex operator which would map to a
``single-trace'' operator which can appear in the space-time OPE; so we see that the leading contribution in the
OPE limit can be associated with such an operator, as we expect. We will verify below that the coefficient also
agrees with our expectations. In any stable background the solutions to this equation obey $j_0 \geq
d/4$.\footnote{There is another solution to \eqref{z saddle} with $j_0 \leq d/4$, which is not relevant for our
discussion since it does not give any solutions to \eqref{j saddle}.} Equation \eqref{j saddle} implies that as
$x\to 0$, the saddle point is at a value of $z$ which also goes to zero (with fixed $\ln|x|/\ln|z|$), but it
remains at a fixed value of $w$ in this limit.

We still have to verify that the saddle has the correct sign of its second derivative and that it is dominant.
The second derivative of the exponent with respect to $w$ is
\begin{equation}
\ln|x| \frac{\d^2}{\d w^2}\left(w\Delta(j_0,h)-w+j_0\right)=-\frac{\ln |x|}{w}\frac{\d j_0}{\d w}.
\end{equation}
The implicit function theorem implies that
\begin{equation}\label{secondyder}-\frac{\ln |x|}{w}\frac{\d j_0}{\d w}=-\frac{\ln
|x|}{w^3}\left(\frac{\d^2 \Delta}{\d j^2}\biggr|_{j_0}\right)^{-1}=\ln |x|\frac{(\d_j\Delta|_{j_0})^3}{\d_j^2
\Delta|_{j_0}}.
\end{equation}
We see that with the same assumption we needed above,
\begin{equation} \label{secondsign}
\d_j^2\Delta|_{j=j_0}<0,
\end{equation}
(assuming also $\d_j \Delta_{j=j_0} < 0$ which is required for the consistency of our computation) the saddle
point is always dominant in the OPE limit $x\to 0$, since the second derivative is proportional to $\ln |x|$.

Now, we drop again all the subleading corrections, and compute the integral (\ref{y integral}) in the saddle
point approximation. The expansion up to quadratic fluctuations around the saddle point $w=w_0$ gives
\begin{gather}\label{saddle y integral}
    I = -i\sqrt{\pi}|x|^{-2j_1-2j_2} \sum_h\int \frac{\sqrt{-\ln |x|} dw}{\sqrt{-w_0\d_j^2\Delta|_{j_0}}}|x|^{2j_0}e^{2\ln|x|\frac{(\d_j\Delta|_{j_0})^3}{2\d_j^2
    \Delta|_{j_0}}(w-w_0)^2}B(j_i,h_i;j_0,h),
\end{gather}
where $j_0$ is now evaluated at the saddle point \eqref{z saddle} so it no longer depends on $w$; the integral
is Gaussian. It is straightforward to compute it and to obtain the final result for the OPE limit of the
$n$-point function (\ref{final-result}),
\begin{gather}\label{final result}
    I = -i\pi\sum_h\frac{|x|^{2j_0-2j_1-2j_2}}{\d_j\Delta(j,h)|_{j=j_0}}B(j_i,h_i;j_0,h),
\end{gather}
using the saddle point relations when necessary. Note that the saddle point is well inside the region where we
trust the various OPE expansions we performed. $j_0$ is such that $\Delta(j_0,h)=1$, and the $x$-dependence of
\eqref{final result} is exactly what we expect for a ``single-trace'' operator of dimension $2j_0$ in the
space-time OPE.

From the space-time point of view, we expect the result to equal what we get by plugging \eqref{stOPE} into the
space-time $n$-point function. This gives the space-time $(n-1)$-point function, times the space-time OPE
coefficient from \eqref{stOPE}, which is equal to the space-time $3$-point function divided by the $2$-point
function of ${\hat O}_{j_0}$. And indeed, the function $B$ includes in its definition above precisely the
(integrated) worldsheet $(n-1)$-point function, times the $3$-point function divided by the $2$-point function
(appearing in the worldsheet OPE). The denominator $\d_j\Delta|_{j_0}$ is a puzzling feature at first sight,
since it is not clear where it comes from in the space-time OPE analysis. It arises from the need to integrate
over the conformal group on the worldsheet when comparing two-point functions on the worldsheet to two-point
functions in space-time \cite{Kutasov:1999xu}. The $j$-independent prefactor $(-i\pi)$ is accounted for in the
same way. Note that exactly the same combination of ``extra factors'' appears also in the analysis
\cite{Maldacena:2001km} of NS-NS string theory on $AdS_3$, and its relation to the integral over the subgroup of
the conformal group which leaves two insertions fixed was explained there.

Thus, after carefully matching the space-time and worldsheet results, we see that (\ref{final result}) is
precisely the expression we would expect for the space-time OPE contributions from ``single-trace'' operators.
Since we got the expression (\ref{final result}) from the leading saddle point approximation, it is natural to
ask what type of corrections, if any, the full integral over the OPE region contributes. This will be analyzed
in \S\ref{secnocorrections}.

It is useful to write the explicit formulae for the saddle point in the special
cases where the spectrum of primaries on the worldsheet is known. In the
case of NS-NS $AdS_3$ (\ref{ads3nsns}) we obtain using our saddle point formulae (\ref{j saddle}), (\ref{z
saddle}), the elegant relation
\begin{equation}\label{rel1}\ln|z| = \frac{k-2}{2j_0-1}
\ln|x|.\end{equation}
When combined with the on-shell relation,
this can also be written as
\begin{equation}\label{rel2}
\ln|z| = \ln|x| \frac{k-2}{\sqrt{4(k-2)(h_0-1)+1}}.\end{equation} The case of weakly curved $AdS_{d+1}\times W$
backgrounds has a similar structure of worldsheet dimensions to the case of NS-NS $AdS_3$, which results in
similar formulae for the saddle point location. Note that in both cases (\ref{secondsign}) is satisfied, so the
existence of the saddle point is guaranteed. Moreover, the saddle point is always at real positive $j_0 \geq
d/4$, which is consistent with the unitarity of the space-time CFT.

Note that our saddle point is really a saddle circle in the integral over $z$. We expect the relevant terms in
the worldsheet OPE to be independent of the angle in the $z$ plane, since the dominant operators which are
relevant for the space-time OPE are necessarily level-matched operators with the same left-moving and
right-moving scaling dimensions. Of course, this does not imply that the correlation functions on the worldsheet
will be independent of the angle in the $z$ plane; these could have an arbitrary dependence on the angle (even
in the small $z$ and small $x$ limits), coming from contributions to the worldsheet OPE from operators with
different left-moving and right-moving scaling dimensions. However, all these contributions vanish upon
integration over the phase of $z$, which is why we could ignore them in our analysis (even though they can give
important contributions to the precise form of the unintegrated worldsheet correlation functions in the OPE
limit). The integral over $z$ is only sensitive to the angle-independent terms in the Fourier decomposition of
the correlation functions, so we lose all the information about the angular dependence.

\subsection{The absence of corrections to the saddle
point}\label{secnocorrections}

From the point of view of the space-time OPE, one can argue that all the perturbative corrections coming from
the expansion around the saddle point must vanish. The reason for this is that the second derivatives around the
saddle point are proportional to $\ln(|x|)$, so any corrections would be powers of $\ln(|x|)$. However, such
corrections (for ``single-trace'' operators) are inconsistent with the space-time OPE, which should only include
powers of $|x|$ (Logarithms can appear when expanding the space-time OPE in some perturbation expansion, but
this is not the case in our computation.).

Let us verify that all these corrections indeed conspire
to cancel. We will verify this directly below, but first we will
use a simpler method. Our starting point is the integral
(\ref{theintegralthree})
\begin{gather}
    |x|^{2j_1+2j_2} I = \sum_h\int d\ln{(|z|)}\int_{C} dje^{2\ln(|z|)(\Delta(j,h)-1)+2\ln(|x|) j}
    B(j_i,h_i;j,h).
\end{gather}
This integral converges (at least in the small $|z|$ region that we are interested in) in all stable
backgrounds, since in such backgrounds all operators obey that $\Delta(j_0,h)=1$ for some $j_0 \geq d/4$
\cite{Breitenlohner:1982jf}, implying that $\Delta(j,h) \geq 1$ on the contour $C$.\footnote{It does not
converge in unstable backgrounds, such as bosonic string theory on $AdS_3$. For the special case of
$j_0=\frac{d}{4}$, the integral does not formally converge, but all our statements are true for this case as
well. } Consequently, we can take a different path from the one we took in our previous analysis, and perform
the integral over $\ln(|z|)$ first, since it is a trivial integral. Assuming that the integral is over the
region $(-\infty,\ln|\epsilon|)$ for some $\epsilon$ with $|\epsilon| < 1$, we obtain
\begin{gather}\label{pole2}
    |x|^{2j_1+2j_2} I = \frac12\sum_h\int_{C} dj\frac{|\epsilon|^{2\Delta(j,h)-2}}{\Delta(j,h)-1}|x|^{2j}
    B(j_i,h_i;j,h).
\end{gather}
We can now try to compute this integral by the saddle point method, as
in the previous subsection. The integral has a saddle point at $j=j_1$
defined by
\begin{equation}
    \d_j\Delta(j,h)|_{j=j_1} =-\frac{\ln |x|}{\ln |\epsilon|}.
\end{equation}
Using our assumption that $\d_j \Delta(j,h)$ is monotonically decreasing along the real axis, and goes to minus
infinity for large positive $j$, we find that as $x\to 0$, $j_1 \to \infty$. However, the integrand of
\eqref{pole2} has a pole at $j=j_0$ defined (as in the previous subsection) by $\Delta(j_0,h)=1$, and in the
$x\to 0$ limit we have that $j_1 \gg j_0$. Thus, as we shift the integral \eqref{pole2} parallel to the real
axis from the contour $C$ to the contour $C'$ given by $j_1+is$ for the saddle point evaluation, we pick up a
contribution from this pole (Generically this is a simple pole since $B$ will not have a pole at the same
place.). Using the Cauchy theorem we then obtain that \eqref{pole2} is equal to
\begin{gather}\label{pole}
    \frac12\sum_h\int_{C'} dj\frac{|\epsilon|^{2\Delta(j,h)-2}}{\Delta(j,h)-1}|x|^{2j}
    B(j_i,h_i;j,h) - i\pi \sum_h\frac{|x|^{2j_0}}{\d_j\Delta(j,h)|_{j_0}}
    B(j_i,h_i;j_0,h),
\end{gather}
ignoring possible additional discrete contributions coming from poles of $B$ which are crossed as the contour is
moved (These will be discussed in the next section.).

Note that the second term of (\ref{pole}) is exactly the expression \eqref{final result} we obtained in the
previous section in the saddle point approximation. On the other hand, the first term may be reliably evaluated
by the saddle point method, and it goes as $|x|^{2j_1}$ as $x \to 0$. Since $j_1 \to \infty$ in the same limit,
this term does not contribute to the OPE expansion (and in any case, since it has explicit $\e$ dependence, we
expect it to cancel in the full computation). This establishes that the leading order result from the saddle
point method for the integral (\ref{theintegralthree}), which we used in the previous subsection, is exact, and
does not obtain any corrections going as powers of $\ln(|x|)$. A very similar method was used to evaluate the
OPE in \cite{Maldacena:2001km}. The disadvantage of treating the integral as we have done in this subsection is
obvious: one loses the information about the double saddle point we discussed in the previous section, and about
having a dominant value of $|z|$ in the computation (In particular, the interesting relations
(\ref{rel1}),(\ref{rel2}) are no longer visible.).\footnote{The fact that the saddle point disappears when we
change the order of integration is possible because the saddle point is away from the original integration
contour, so it does not have to appear in both orders of integration even though the original integral
converges. Indeed, when we shift the contour to $j_0+is$ so that the saddle point is on the contour of
integration, the integrals no longer commute.} However, the final result is the same.

We can also verify directly that there are no corrections to the saddle point. After the change of variables to
$w=\ln(|z|)/\ln(|x|)$, the integral is proportional to
\begin{gather}
    \ln|x|\sum_h\int dw\int_{C} dje^{2\ln |x|\left(w\Delta(j,h)-w+j\right)}
    B(j_i,h_i;j,h).
\end{gather}
Expanding this systematically around the saddle point and
diagonalizing the quadratic part in the exponent, after the dust
settles, one encounters integrals of the form
\begin{gather}
\int ds \int dte^{-s^2-t^2}(s+it)^nt^m,
\end{gather}
with $n>m$. Such integrals vanish, as can be seen by using the $SO(2)$ symmetry in the $s,t$ plane. The domain
of integration is not precisely $SO(2)$ invariant, but the non-invariant contributions are far from the saddle
point and vanish rapidly as $x\to 0$, as in the previous computation. In this way, one can directly see that the
saddle point receives no corrections which are inconsistent with the space-time OPE.

\section{Discrete
contributions}\label{discretecontributionssection}

In this section we will analyze the additional discrete contributions to the computations above which come from
poles in the integrand (\ref{theintegralthree}) as a function of $j$. Such contributions can arise both when the
poles cross the integration contour during the analytic continuation in $j_1$ and $j_2$, and when they cross the
contour as we shift it from $j=d/4+is$ to $j=j_0+is$. In the special case of $AdS_3$ with NS-NS flux, the pole
structure of $B$ was analyzed in \cite{Maldacena:2001km}. It was found that there are two relevant families of
poles in the structure function $B$, which appear at
\begin{equation}\label{poles1}
    j=j_1+j_2+n \ ,\ n\in \{0,1,2,\cdots\},
\end{equation}
and at
\begin{equation}\label{poles2}
    j=|j_1-j_2|-n \ ,\ n\in \{0,1,2,\cdots\}.
\end{equation}
Recall that $B$ includes a worldsheet $3$-point function (which is the same
as the space-time $3$-point function) of operators with dimensions $2j$, $2j_1$ and
$2j_2$. In the case \eqref{poles1} the pole is related to the mixing of (a descendant of) the
product of the latter two operators with ${\hat O}_j$, and the case \eqref{poles2} is
related to this by a permutation of the operators. Thus, we expect
both types of poles to be completely general, and to appear in any example
of the $AdS_{d+1}/CFT_d$ correspondence.

From the point of view of our computation of the space-time OPE, the naive interpretation of a contribution from
the first family of poles (\ref{poles1}) is as a ``double-trace'' operator of dimension $2j$ (related to ${\hat
O}_{j_1} {\hat O}_{j_2}$) appearing in the OPE. On the other hand, the second family of poles \eqref{poles2} has
no straightforward interpretation in space-time since, generically, there are no physical space-time operators
with the appropriate dimensions $2j$. We will first discuss the family of poles (\ref{poles1}), and later turn to
(\ref{poles2}) which behave a little differently.

Let $h$ be fixed and consider the contribution coming from crossing a pole of the form (\ref{poles1}). 
These poles are not crossed during the analytic continuation in $j_1,j_2$, but they may be
crossed when we shift our contour.
We begin with our general result for the form of the
contribution in the OPE region
\begin{gather}\label{theintegralthree1}
   I_1 = |x|^{-2j_1-2j_2} \int_{-\infty}^{\ln|\e|} d\ln{|z|}\int_{C} dj|z|^{2(\Delta(j,h)-1)}|x|^{2j} B(j_i,h_i;j,h),
\end{gather}
with $C=\frac d4+is$. 
As in the previous section, one way to compute this is to first perform the $z$ integral, leading to
\begin{gather}\label{theintegralthree2}
   I_1 = |x|^{-2j_1-2j_2} \int_{C} dj\frac{|\e|^{2(\Delta(j,h)-1)}}{2(\Delta(j,h)-1)}|x|^{2j} B(j_i,h_i;j,h).
\end{gather}
For simplicity, we assume that the poles of the series (\ref{poles1}) are not coincident with physical operators
satisfying $\Delta=1$. In this case, the integrand of (\ref{theintegralthree2}) has isolated single poles.
Moreover, the integrand has a saddle point with $\d_j\Delta\sim -\ln |x|/\ln|\e| $ which is very far away on the
positive $j$ axis, and does not contribute to the OPE limit. Hence, we are able to evaluate the integral by a
contour argument, with the result
\begin{gather}\label{theintegralthree4}
   I_1 = -i\pi|x|^{-2j_1-2j_2}\Biggl(\frac{|x|^{2j_0}}{\d_j\Delta(j,h)|_{j_0}}
    B(j_i,h_i;j_0,h)+\cr+\sum_n\frac{|\e|^{2(\Delta(j_1+j_2+n,h)-1)}}{\Delta(j_1+j_2+n,h)-1}|x|^{2(j_1+j_2+n)}
    Res_{j=j_1+j_2+n}[B(j_i,h_i;j,h)]\Biggr).
\end{gather}
Note that the first term is our usual on-shell contribution, and the sum over $n$ is over the
contributions from poles of the series (\ref{poles1}). To understand how to deal with these contributions, 
we can consider the opposite order of integration, where we first perform the $j$ integral in
\eqref{theintegralthree1}.

The integrand in this case has a simple pole at $j=j_1+j_2+n$ coming from $B(j_i,h_i;j,h)$. In addition, there
is a saddle point of the $j$ integrand at $\d_j\Delta(j,h)=-\frac{\ln|x|}{\ln|z|}$. Denote the solution of this
equation by $j_0\left(\frac{\ln|x|}{\ln|z|}\right)$. Since it is a monotonically increasing function, if we
shift the contour $C$ to the contour $j_0\left(\frac{\ln|x|}{\ln|z|}\right)+is$, we collect the poles of $B$
only if $|z|>|z|_{cr}$ where
$$\d_j\Delta(j,h)|_{j=j_1+j_2+n}=-\frac{\ln|x|}{\ln(|z|_{cr})}.$$
Thus, the discrete contribution takes the form
\begin{gather}\label{eqdisc}
    \int_{\ln|z|_{cr}}^{\ln|\e|} d\ln{|z|}e^{2\ln |z|(\Delta(j_1+j_2+n,h)-1)+2\ln |x| (j_1+j_2+n)}
    Res_{j=j_1+j_2+n}[B(j_i,h_i;j,h)].
\end{gather}
As long as $|z|$ satisfies $|z|>|z|_{cr}$ and is away from $|z|_{cr}$ this is indeed reliable, and the saddle
point evaluation of the $j$ integral is legal. However, when $|z|$ is sufficiently close to $|z|_{cr}$ the
saddle point evaluation of the $j$ integral breaks down, and one can verify that from this region we obtain a
contribution like the one in (\ref{eqdisc}) but completing the region of integration to $(-\infty,\ln|\e|)$.
Thus, in this order of integration the final result takes the following (equivalent to
(\ref{theintegralthree4})) form\footnote{Note that, in particular, (\ref{z integral}) is not reliable near poles
of the structure function $B$ since there is no faithful saddle point of the $j$ integral in this regime. Thus,
(\ref{z integral}) is only correct near the saddle point, and the other regions actually combine to give the
more general result (\ref{theintegralthree3}).}
\begin{gather}\label{theintegralthree3}
   I_1 = -i\pi|x|^{-2j_1-2j_2}\Biggl(\frac{|x|^{2j_0}}{\d_j\Delta(j,h)|_{j_0}}
    B(j_i,h_i;j_0,h)+\cr+2\sum_n\int_{-\infty}^{\ln|\e|} d\ln{|z|}e^{2\ln |z|(\Delta(j_1+j_2+n,h)-1)+2\ln |x| (j_1+j_2+n)}
    Res_{j=j_1+j_2+n}[B(j_i,h_i;j,h)]\Biggr).
\end{gather}

In this way of writing the discrete contributions to $I_1$, it is clear that they are analogous to a well known
behavior in the OPE limit of string theory in flat space, so let us  recall some basic features of it. Consider
the S-matrix for four tachyons
\begin{equation} \langle e^{ik_1X}e^{ik_2X}e^{ik_3X}e^{ik_4X} \rangle, \end{equation}
with all tachyons on-shell. Suppose we choose a gauge in which the vertices are localized
at $z,0,1,\infty$ (in this order), and
consider the contributions from the OPE limit $z\rightarrow 0$. In this limit the result is dominated by the leading term in the OPE
which takes the form
\begin{equation} e^{ik_1X}(z)e^{ik_2X}(0)\sim |z|^{\alpha'k_1\cdot k_2}e^{i(k_1+k_2)X}(0)+\cdots. \end{equation}
Thus, the contribution to the correlation function from the small $z$ region is
\begin{equation} 2\pi \int_{|z|<\epsilon} d|z|  |z|^{\alpha'k_1\cdot k_2+1}
\langle e^{i(k_1+k_2)X}(0)e^{ik_3X}(1)e^{ik_4X}(\infty)\rangle. \end{equation}
There are two possible cases; one is that $e^{i(k_1+k_2)X}(0)$ is off-shell
and the other case is that it is on-shell.
Note also that
\begin{equation} \alpha'k_1\cdot k_2+1=\frac{\alpha'}2(k_1+k_2)^2-3. \end{equation}
The integral is convergent when
\begin{equation} \frac{\alpha'}2(k_1+k_2)^2-3>-1 \qquad \Rightarrow \qquad \frac{\alpha'}2(k_1+k_2)^2-2>0. \end{equation}
The intermediate state is on-shell exactly when the equality $\frac{\alpha'}2(k_1+k_2)^2-2=0$ is satisfied.
There is a pole in the space-time scattering amplitude when the intermediate state is on-shell. For other
values, the contribution to the space-time amplitude is obviously small, and goes to zero when
$\epsilon\rightarrow 0$. This is correct even though there is divergence in the integral,
since we can use an
analytic continuation which states that as a distribution
\begin{equation} Res_{j=-2}[|z|^j] = 2\pi \delta^2(z,\bar z). \end{equation}
This results in a pole in the space-time scattering amplitudes when the intermediate state is on-shell,
and when the intermediate state is off-shell there is no contribution to the space-time scattering amplitude from
the small $|z|$ region.

The discrete contributions in $AdS$ backgrounds coming from (\ref{poles1}) are analogous to those in flat space,
and the structure of divergences is very similar. The generic situation is that the poles (\ref{poles1}) are not
coincident with $j_0$, so the form of the contribution is indeed (\ref{theintegralthree3}).\footnote{There is an
interesting special case where some discrete contribution of the type (\ref{poles1}) equals $j_0$, so that we
have a double pole in the integral over $j$ (if we first do the $z$ integral). This case should be treated
separately since both terms in (\ref{theintegralthree2}) or in (\ref{theintegralthree3}) are infinite. However,
their sum is finite, and can easily be evaluated from the double pole. The result contains a $\ln(|x|)$ factor,
which may seem problematic from the point of view of an OPE in a conformal theory. However, in this case there
is a ``double-trace'' operator of dimension $2j_0$ (in the limit $g_s\to 0$) appearing in the OPE. This
``double-trace'' operator can have an anomalous dimension (of order $g_s^2$), and this can give rise to a
logarithm in the leading order OPE computation, as in \cite{SUGRA-GAUGE OPE} (it is a subleading correction to
the double sphere diagram which usually gives the ``double-trace'' contribution to the OPE, so it is of the same
order as the ``single-trace'' contributions). Contributions to the coefficient of this logarithm may arise also
from the $|z|>|\e|$ region of the worldsheet theory, which may or may not cancel the contributions from the
worldsheet OPE region.} Using the same logic as in the previous paragraphs, we conclude that there are no
contributions from the poles (\ref{poles1}) to the space-time OPE from the $|z|<|\epsilon|$ region. Even though
naively we might have thought that contributions from \eqref{poles1} would be related to ``double-trace'' terms
in the space-time OPE, we see that this does not generically happen (in the worldsheet OPE region), in agreement
with the general expectation that such contributions should not arise locally on the worldsheet.

Let us now discuss the other family of poles, (\ref{poles2}). The basic difference from (\ref{poles1}) is that
during the analytic continuation of $j_1$ and $j_2$ some of them may cross the contour $C$ (The ones who do not
cross $C$ are not crossed also later, so they do not contribute anything.). This adds to
(\ref{theintegralthree1}) a discrete contribution (which is just an integral over $z$). The evaluation of
(\ref{theintegralthree1}) proceeds as before. For convenience, we analyze this by doing the $z$ integral first.
The single pole at $j=|j_1-j_2|-n$ contributes exactly the opposite of the contribution from crossing the
contour $C$. Hence, this family of poles does not give any contributions to $I_1$.\footnote{Similar arguments
imply that any other poles of $B$ also do not contribute, except perhaps when they coincide with $j_0$.}

We can split the complete integrated $n$-point worldsheet amplitude
\begin{equation}\label{discretesect}\int
d^2z\left<O_{1}(x;z,\bar{z})\bar ccO_{2}(0;0)X\right>
\end{equation}
to a contribution $I_1$ from $|z|<|\epsilon|$ which we analyzed above
and a contribution $I_2$ from $|z|>|\epsilon|$ which is not known
analytically except for some very special cases.
The integral $I_1$ was argued, generically, not to have any finite contributions to the space-time OPE except
for the ``single-trace'' contributions coming from the saddle point,
\begin{gather}
    I_1 \simeq -i\pi\sum_h\frac{|x|^{2j_0-2j_1-2j_2}}{\d_j\Delta(j,h)|_{j_0}}
    B(j_i,h_i;j_0,h).
\end{gather}
The integral $I_2$ is not known in general, but in the special case of 4-point functions on $AdS_3$
\cite{Maldacena:2001km} it was argued that the integrand in this region could also be written as an integral of
$j$ over the contour $C$, and it was convenient to shift this integral to a different contour
$C''=\frac{k-1}{2}+is$. The contribution from the integral over this contour was shown to lead to contributions
to the space-time OPE from intermediate states which were identified with winding number $w=1$ strings. On the
way, some poles of the integrand of $I_2$ were crossed, and these poles were claimed to be associated to
``double-trace'' operators in the space-time OPE. It would be interesting to understand more generally how to
see the ``double-trace'' contributions to the space-time OPE.

\section{A comparison to Gopakumar's proposed dual of free gauge
theories}

The results described in the previous sections may have various applications. Here we discuss one application,
which is a test of the proposal of R.~Gopakumar for the string dual of free large $N$ gauge theories
\cite{Gopakumar:2005fx}.\footnote{For a review see \cite{Aharony:2006th}. This proposal can be used for
perturbative gauge theories as well.} Gopakumar proposed a specific string theory dual for these theories, but
it is not presented in the form of a local worldsheet theory, so we cannot directly use our arguments from the
previous sections. We wish to check whether in the space-time OPE limit, this proposed dual exhibits a dominant
saddle point with $\ln|z|$ scaling as $\ln|x|$, as implied by our general analysis of \S \ref{main}.

For planar diagrams, the prescription of \cite{Gopakumar:2005fx} involves first rewriting each free field theory
Feynman diagram contributing to an $n$-point correlation function as an integral over the Schwinger parameters
$\sigma_{ij}$ of the propagators connecting the $i$'th and $j$'th operators. For each set of propagators which
connect the same two vertices with contractible color flow lines we use a single Schwinger
parameter.\footnote{The original prescription of \cite{Gopakumar:2005fx} involved Schwinger parameters in
momentum space, but it will be more convenient for us to use Schwinger parameters in position space. These arise
by replacing the four dimensional scalar propagator $1/|x|^2$ by $\int_0^{\infty} d\sigma e^{-\sigma |x|^2}$,
and they are the inverses of the momentum-space Schwinger parameters. When joining adjacent propagators
together, the $\sigma$'s simply add together.} Then, one maps the space of these Schwinger parameters to the
``decorated moduli space'' ${\mathcal M}_{0,n}\times \mathbb R_+^n$: the moduli space of a sphere with
$n$ marked points, together with a positive number $p_i$ associated with each marked point. As described in
detail in \cite{Gopakumar:2005fx}, this mapping uses the properties of Strebel differentials. After integrating
over the $p_i$, this procedure gives a specific worldsheet $n$-point correlation function associated with the
original correlator, which, by construction, reproduces the correct $n$-point space-time correlation function
(upon integration over the worldsheet moduli).

The Schwinger representation (in position space) of a generic tree-level amplitude of a free gauge theory
(specializing to the case of a four dimensional correlation function of single-trace operators made from adjoint
scalar fields) takes the form
\begin{equation}\label{treegeneric}
\int d\sigma_{12}\sigma_{12}^{J_{12}-1}\int\prod_{\{k,l\}\neq
\{1,2\}} [d\sigma_{kl} \sigma_{kl}^{J_{kl}-1}]
e^{-\sigma_{12}x_{12}^2-\sum_{\{k,l\}\neq\{1,2\}}\sigma_{kl}x_{kl}^2},
\end{equation}
where $J_{kl}$ are the numbers of propagators connecting the $k$'th and $l$'th vertices (separated by a distance
$x_{kl}$), and the product and sum over $\{k,l\}$ go over all pairs of vertices with $J_{kl} \geq 1$. We will be
interested in the limit $x_{12} \to 0$ where the first two operators come together; we assume $J_{12} > 0$ since
otherwise no single-trace operators contribute to this OPE limit. Defining new variables $T_{kl}\equiv
\sigma_{kl}/\sigma_{12}$ for $\{k,l\}\neq \{1,2\}$, we can integrate over $\sigma_{12}$ which plays the role of
the overall scale. We remain with an integral of the form (up to a constant)
\begin{equation} \label{tklints}
\int_{\mathbb{R}^{m}_+}\prod_{kl}dT_{kl}\frac{\prod_{kl}T_{kl}^{J_{kl}-1}}{(x_{12}^2+\sum_{kl}T_{kl}x_{kl}^2)^M},
\end{equation}
where $m$ is the number of lines in the reduced Feynman diagram minus one (for a generic $n$-point planar
diagram, $m=3n-7$), and $M\equiv J_{12}+\sum_{kl} J_{kl}$. Next, we transform to spherical coordinates in the
space of $\{T_{kl}\}$. The integral over the radial coordinate $r$ in this space depends only on the ratio
$r/x_{12}^2$, up to an overall factor (independent of $r$). Hence, the main contribution to the integral (which
is non-negative and convergent) in the $x_{12}\to 0$ limit comes from the region where $r$ is of order
$x_{12}^2$. However, there is no dominant saddle point in this region (say, working in the variable
$r/x_{12}^2$).

This reasoning establishes that the integral is dominated by a region in which the ratio of the Schwinger
parameter $\sigma_{12}$ to the other $\sigma_{kl}$'s scales as $1/x_{12}^2$. Next, we should study how this
region maps into the moduli space of Riemann surfaces by the mapping of \cite{Gopakumar:2005fx}. This was
studied in \cite{Gopakumar:2005fx} (and later corroborated in explicit examples\footnote{One can also prove this
result in general with less assumptions than were used in \cite{Gopakumar:2005fx}.}
\cite{Aharony:2006th,David:2006qc,Aharony:2007fs}), where it was shown that this region maps to the region of $z
\to 0$ (where $z$ is the distance between the worldsheet positions of the first two vertex operators), with
\begin{equation} \label{Strebel scaling1} |z| \propto r.\end{equation}
Combining this with the scaling relation $r/x_{12}^2\sim 1$, we find that the dominant contribution to
the worldsheet correlation functions in the OPE limit comes from a region in which \cite{Gopakumar:2005fx}
\begin{equation} \label{Strebel scaling2} |z| \propto x_{12}^2.\end{equation}

This is somewhat similar to the results of \S \ref{main}, which indeed predict such power law scaling relations.
Assuming that the origin of this scaling is similar to that in the discussion of \S \ref{main}, we can rewrite
(\ref{Strebel scaling2}) as
\begin{equation} \label{freescaling} \frac{\ln(|z|)}{\ln(|x_{12}|)}=2,\end{equation}
and find using \eqref{j saddle} that the worldsheet scaling dimensions of the corresponding worldsheet theory must obey
\begin{equation} \partial_j \Delta(j,h)|_{j=j_0} = -\frac{1}{2}, \end{equation}
for any operator (which becomes physical at $j=j_0$).

However, such a conclusion is too naive since the detailed behavior of the integral (\ref{tklints}) (interpreted
as an integral over $|z|$ instead of over $r$) is quite different from the behavior we found in \S \ref{main}.
In particular, there is not really a saddle point in this integral, while in \S \ref{main} we proved that the
existence of a continuum of normalizable vertex operators, with a well-behaved function $\Delta(j,h)$
determining their worldsheet scaling dimensions, leads to a dominant saddle point.\footnote{In
fact, the analysis of the saddle
 point in the integral over $j$ in \S \ref{main}
implies that the integrand of the integral over $|z|$ in the OPE limit $z,x\to 0$, must take the form
$|x|^{f\left(\frac{\ln|x|}{\ln|z|}\right)}\times subleading$. This is not obeyed by \eqref{tklints}.}

Thus, the formalism of \cite{Gopakumar:2005fx} leads to results which are similar, but not identical, to our
general expectations from \S \ref{main}. This suggests that this formalism may involve a somewhat singular
worldsheet theory, as expected on general grounds for the dual of free gauge theories, and as suggested by
various other properties of this formalism \cite{Aharony:2006th}. Some possibilities are :
\begin{itemize}
\item The integral over $j$ may not have a saddle point in the limit of free gauge theories; for
instance, the function $\Delta(j,h)$ may not even be analytic in this limit. Even with
non-analytic functions we have been unable to obtain integrands of the form \eqref{tklints},
but we cannot prove that this is impossible.

\item There may not be a continuum of normalizable vertex operators on the worldsheet in the limit
of free gauge theories. Note that our arguments for the existence and form of such a continuum involved using
the space-time conformal symmetry, while this symmetry is not \cite{Aharony:2006th} realized locally in the
formalism of \cite{Gopakumar:2005fx}. It would be interesting to find arguments for or against such a continuum
in the string dual of free gauge theories.

\item The formalism of \cite{Gopakumar:2005fx} may not give rise to standard local worldsheet theories as we assumed in \S \ref{main}.

\end{itemize}

\section*{Acknowledgments}

We would like to thank D.~Kutasov for collaboration in the early stages of this project, and for many incisive
comments and insights. We also thank M.~Berkooz, J.~David, R.~Gopakumar, S.~Razamat, and A.~Schwimmer for useful
discussions. ZK would like to thank Stanford University and SLAC for hospitality during early stages of this
work. OA would like to thank the University of Chicago, Stanford University, SLAC and the Isaac Newton Institute
for Mathematical Sciences for hospitality during the course of this work. This work was supported in part by the
Israel-U.S. Binational Science Foundation, by a center of excellence supported by the Israel Science Foundation
(grant number 1468/06), by a grant (DIP H52) of the German Israel Project Cooperation, by the European network
MRTN-CT-2004-512194, and by a grant from G.I.F., the German-Israeli Foundation for Scientific Research and
Development.

\end{document}